\documentclass[preprint,aps,prc,showpacs,nofootinbib]{revtex4}

\usepackage{epsfig}
\usepackage{amsmath,graphicx}

\newcommand\ba{\begin{eqnarray}}
\newcommand\ea{\end{eqnarray}}
\newcommand\nn{\nonumber}

\begin{document}
\title{Muon multiplicity at high energy proton-nuclei collisions}

\author{E. A. Kuraev, S. Bakmaev, V. Bytev, E. Kokoulina}
\affiliation{\it JINR-BLTP, LPP, 141980 Dubna, Moscow region, Russian
Federation}

\date{\today}

\begin{abstract}
Estimation of multiplicity of muons and pions production at high energy
proton-nuclei collisions is given. Both QED and QCD contributions are
considered for peripheral kinematics of muon pair and $\sigma$-meson
production, keeping in mind it's final conversion to muons. An attempt
to explain the excess of positive charged muons compared to negative
one in cosmic muon showers is given.

We derive the dependence of cross-section of $n$ pairs as a function of $n$
at large n as $d^n(n!n^2)^{-1}$.
\end{abstract}

\maketitle

\section{Introduction}

The essential impact parameter in pair of charged particle (muons
or pions) creation process at collision of high energy particle
(proton or nuclei from cosmic ray) is of the order
$\frac{1}{m_\pi}\sim\frac{1}{m_\mu}=\frac{1}{m}\sim 2 fm$. Such a
particle penetrating through the Earth atmosphere at least once
will collide the nuclei of gas (nitrogen or oxygen). It
corresponds to direct collision with small orbital momentum or
impact factor $\rho<1 fm$. Besides the pair can be created in
peripheral collisions which are responsible for large values of
orbital momentum $\rho>1 fm$. It is interesting to evaluate the
number of high energy muons which reach the surface of the Earth.
In particular the problem of explaining the exceed of positive
muons or rather high energies over the negative ones discussed
first in \cite{Frazer}. This number is the sum of the numbers of
created muons and pions, as well all the pions due to main decay
mode will turn to muons. As for creation of light lepton pairs
(electron and positron) their fate is to create the showers which
do not produce muons. The probability of creation of heavy lepton
pairs such as tau-meson is suppressed compared with muons one by
factor $(m/m_\tau)^2\sim 10^{-3}$ and will not be considered
below.

\section{QED mechanism of muons production}

Cross section of production $\mu^+\mu^-$ pair at high energy
proton-nuclei collision have a form \ba \label{ll}
\sigma^{QED}_{(1)}=\frac{28(Z\alpha^2)^2}{27\pi
m^2}[L^3-2.2L^2+3.8L-1.6-3L^2f(Z\alpha)], \quad
f(t)=t^2\sum\nolimits_1^\infty\frac{1}{n(n^2+t^2)} \ea with
$L=\ln(2\gamma), \gamma=\frac{E}{M}, E, M$- energy and mass of
proton in laboratory frame, $Z$ is the nuclei charge and $m$ is
the muon mass. The first 4 term in square bracket (\ref{ll}) was
obtained by Racah \cite{Racah}, the last one, which takes in to
account many photon exchanges was obtained by Bethe and Maximon
\cite{Bethe}.

For $L=15$, $Z=7$ we have $\sigma^{QED}_{(1)}\approx 5\times
10^{-30}cm^2$. We estimate the number of muon pairs created by a
single proton as $N_\mu^{QED}\approx
M_P^2\sigma^{QED}_{(1)}\approx 10^{-2}$.

Production of several pairs of charged particles contribution to
the cross section is not associated with $L^3$ enhancement. It was
shown in \cite{Bartosh}  that the distribution on the impact
parameter of probability  of $n$ lepton pairs production in
ion-ion collisions have a Poisson form:
\ba
\frac{d\sigma_n}{d^2\rho}=P_n(A_l),\qquad
P_n(z)=\frac{z^n}{n!}e^{-z}, \ea with \ba
A_l=\frac{28(Z\alpha^2)^2}{9\pi^2\rho^2 m^2}\Phi(\rho,\gamma),
\ea
$\rho$-impact parameter-the distance in transversal direction
between proton and nuclei, $\Phi(\rho,\gamma)$ will be given
below.

This result can be generalized for the case of production of
arbitrary number of lepton $n_l$ and pion $n_\pi$ pairs with fixed
total number $n=n_l+n_\pi$: \ba
\frac{d\sigma^{QED}_n}{d^2\rho}=\frac{1}{n!}(A_l+A_\pi)^n
e^{-A_l-A_\pi}, n=2,3,... . \ea with \cite{BaierFadin}
$A_\pi\approx A_l/14$. This estimation is based on the ratio of
leading cross sections in Born approximation and the relation
$1/m^2_\pi\approx 1/(2m^2)$. This results in replacement
$A_l+A_{\pi}$ by $A_l $ and the coefficient $28/9$ in expression
for $A_l$ by $10/3$.

The quantity $\Phi(\rho,\gamma)$ have a form \cite{HKS07}: \ba
\Phi(x,\gamma)&=&\ln\frac{\gamma^2}{x}\ln\frac{x}{x_0}+\frac{1}{4}\ln^2\frac{x}{x_0},\quad
x_0<x<\gamma^2, \nn \\
\Phi(x,\gamma)&=&(\ln\gamma^2-\frac{1}{2}(\ln x_0+\ln x))^2,\quad
\gamma^2<x<\frac{\gamma^4}{x_0x},
\nn \\
 x&=&(m\rho)^2,\quad  x_0=(mR)^2,
\ea with $R$-nuclei radii.

The cross section of creation of $n$ pairs in the single collision
will be \ba \sigma^{QED}_n(\gamma)=\frac{\pi B^n}{m^2
n!}I_n(\gamma, B),\quad n>1, \ea with
$B=\frac{10}{3\pi^2}L(Z\alpha^2)^2$ and \ba
I_n=\int\limits_{x_0}^\infty\frac{dx}{x^n}\Phi^n
e^{-\frac{B}{x}\Phi}. \ea

For $L=15, Z=7$ we have $B\approx 7.05\times10^{-7}$.
 The Quantity
$I_n(\gamma,A)$ for $x_0=(R(fm))^2/4\approx 1$ can be approximated
as $I_2\approx 8L^2; I_3\approx 3L^3; I_4\approx 1,58 L^4, ...$.

For the case $n=1$ we must take into account the unitary corrections to the cross
section in Born approximation which will be considered below.
\ba
\sigma^{QED}_1=\sigma^{QED}_B+ \frac{B}{m^2}\int\frac{d^2\rho}{\rho^2}\Phi
(e^{\frac{-B\Phi}{\rho^2}}-1)\mid_{\rho>R}.
\ea

\section{Pomeron exchange mechanism of muons production}

In a simplified Pomeron model \cite{GSFK79} the exchange by two reggeized gluons in the
scattering channel is associated with Pomeron Regge pole exchange.
Pomeron-proton coupling is described by
the effective vertex
\ba
\Phi_P(\vec{l}_1,\vec{l}_2)^{\lambda_1\lambda_2}=-\frac{12\pi^2}{N_c}F_P(\vec{l}_1,\vec{l}_2),
\quad N_c=3;\nn \\
F_p(\vec{l}_1,\vec{l}_2)=\frac{(-3\vec{l}_1\vec{l}_2)C^2}{[C^2+(\vec{l}_1+\vec{l}_2)^2]
[C^2+\vec{l}_1^2+\vec{l}_2^2-\vec{l}_1\vec{l}_2]}, \quad C=m_\rho/2;
\ea
$\vec{l}_1,\vec{l}_2=\vec{l}-\vec{l}_1$-are two-dimensional vectors of gluons.

The vertex of emission of $\sigma$-meson in collision of two Pomerons can be obtained
using the $RRPP$ effective vertex obtained in the paper \cite{ACKL05}.
Really one can consider only the kinematic region when the 4-momenta of the gluons
almost equal $p_1=p_2\approx(q_1+q_2)/2=p/2,\quad p^2=M^2_\sigma$ with $q_{1,2}$-the momenta
of reggeized gluons, $M_\sigma$-is the $\sigma$-meson mass. Projecting this vertex on
the colorless and spin-zero state we obtain:
\ba
\frac{1}{\sqrt{N_c^2-1}}\delta^{a_1a_2}g^{\nu_1\nu_2}\Gamma^{-\nu_1\nu_2+}_{ca_1a_2d}(q_1;p_1,p_2;q_2)=
\frac{-4\pi\alpha_s}{\sqrt{N_c^2-1}}N_c\delta^{cd}I(q_1,q_2),
\ea
with
\ba
I(\vec{q}_1,\vec{q}_2)=12-\frac{8}{M^2_\sigma+2\vec{q}_1^2+2\vec{q}_2^2}\left[10(M^2_\sigma+
\vec{q}_1^2+\vec{q}_2^2)-\frac{5}{2}(M^2_\sigma+(\vec{q}_1+\vec{q}_2)^2)+\right.\nn \\
\left.\frac{16\vec{q}_1^2\vec{q}_2^2}{M^2_\sigma+(\vec{q}_1+\vec{q}_2)^2}\right].
\ea
Matrix element of process of single $\sigma$-meson production in
proton-proton collisions have a form
\ba
M^{pp\to pp\sigma}(l,p)=is A_1f(l,p)F(\Delta)\frac{2^7\alpha_s^3\pi^2N_c}{\sqrt{N_c^2-1}},
\ea
with $A_1=A-A^{1/3}$ is the number of nucleons inside the nuclei with atomic number $A$
which interact with the high energy proton by Pomeron exchange,
\ba
f(l,p)=\int\frac{d^2l_1 C^4}{2\pi\vec{l}_1^2(\vec{l}-\vec{l}_1)^2(\vec{p}-\vec{l}_1)^2}
F_P(l_1,l-l_1)F_P(l_1-l,p-l_1)I(\vec{l}_1,\vec{p}-\vec{l}_1),
\ea
and form-factor of the two gluon bound state $F(\Delta)=[a^2\Delta^2+1]^{-2}$,
the relative momentum of gluons $\Delta=|\vec{p}_1-\vec{p}_2|/2$ and $a$ is the
size of two gluons bound state $a\approx 1 fm$.
Performing the phase volume of the final 4-particle state as
\ba
d\Gamma_4=\frac{d^3p_a'}{2E_a}\frac{d^3p_b'}{2E_b}\frac{d^3p_1}{2\omega_1}
\frac{d^3p_2}{2\omega_2}(2\pi)^{-8}\delta^4(P_a+P_b-P_a'-P_b'-p_1-p_2)= \nn \\
\frac{d^2l_1d^2p}{(2\pi)^22s}L\frac{d^3\Delta}{(2\pi)^3M_\sigma}(2\pi)^{-3}
\ea
After integration over $\Delta$
\ba
\int\frac{d^3\Delta}{(2\pi)^3M_\sigma}F^2(\Delta)=\frac{1}{32\pi M_\sigma a^3}
=\frac{M^3_p}{32\pi5^3(a(fm))^3M_\sigma}.
\ea
The cross section of sigma meson production in Born approximation can be written in form:
\ba
\sigma^p_{01}=\frac{9A_1^2\alpha^6_sL}{4000}
\frac{M^3_p}{M_\sigma m^4_\rho}J,\\
J=\int\frac{d^2ld^2p}{(2\pi)^2C^4}f^2(l,p)\approx 7.4*10^3.
\ea
Numerical estimation gives for $M_\sigma\approx C\approx 400MeV, J\approx 1$.

Screening effect are taken into account below(see (\ref{eq24}), n=1).
Consider now process of several $\sigma$-meson production at proton-nuclei
peripheral collisions.

At large impact parameters limit proton interact with the whole gluon field
of the nuclei coherently. So main contribution arises from many Pomeron exchanges
mechanism (compare with the "chain" mechanism essential at BFKL equation formation).
The relevant Feynman diagrams-are the s-channel iteration of Pomeron exchange.
Let us consider three kinds of iteration blocks. One is the pure Pomeron exchanges,
the second one-Pomeron exchange with the vertex of emission of external $\sigma$-
meson insertion. The third one-so called "screening block"-two blocks of second type
with common virtual $\sigma$-meson Green function. Contribution of the last one is
associated with boost logarithm-in quite analogy with the problem of several leptonic
pairs production at ions collision considered in paper of one of us \cite{Bartosh}.
In the similar way the closed expression for the summed on numbers of ladders of
the first and the third type can be obtained using the relation
\ba
\int\Pi_1^n\frac{d^2k_i}{(2\pi)^2}=\int\Pi_1^{n+1}\frac{d^2k_i}{(2\pi)^2}\int d^2\rho \nn \\
exp(i\vec{\rho}\sum_1^{n+1}(\vec{k}_i-\vec{q}))=\int d^2\rho e^{-i\vec{q}\vec{\rho}}\Pi_1^{n+1}\frac{d^2k_i}{(2\pi)^2}e^{i\vec{k}_i\vec{\rho}}.
\ea
Omitting the details which are similar to ones given in \cite{Bartosh} we arrive to the
expression for the cross section
\ba \label{eq24}
\sigma^P_n=\frac{A^2_1}{m_\rho^2}\int \frac{C^2d^2\rho}{2\pi} \frac{z^n}{n!}e^{-z},\quad n=1,2,3,...
\ea
with
\ba
z(\rho)=\frac{9L\alpha_s^6 M_p^3 D}{4*10^3(a(fm))^3M_\sigma m_\rho^2},
\quad D=\int\frac{d^2ld^2p}{(2\pi)^2C^4}f^2(l,p)e^{i\vec{p}\vec{\rho}}
\approx e^{-C\rho}J.
\ea

Let us estimate the behavior of $\sigma^P_n$ at large $n$. Keeping in mind
the numerical smallness of $z$ in (20) we have
\ba
\frac{\sigma^P_n}{\sigma^P_1}\approx \frac{d^n}{n^2 n!},d\approx 0.000425 L, a\approx 1 fm.
\ea

\section{Excess of positively charged muons produced in cosmic rate interaction with the Earth}

Let us apply the results obtained above to the problem of explain of exceeding
number of positive charged muon compared the negative charged muons
created by cosmic ray (preferentially consisting from protons).

Rather rough estimation leads to conclusion that a high energy proton crossing
the atmosphere normally to Earth surface averagely in $1.2$ events collide with
nuclei of a gas (nitrogen or oxygen). It excite the nucleons which are on it's
way through the nuclei and interact peripherally with the other nucleons.
Number of the first ones is $N_d\approx A^{1/3}$, number of others is $N_p\approx
A_1=A-N_d$. Resonances decay product is mainly positive charged pions and the
last ones reach the Earth surface as positive charged muons.
We believe  that the direct type collisions corresponds to values of impact
parameter of order of $1fm$-the proton radius. For peripheral collisions we
choice $\rho>1fm$.

The peripheral interactions produce equal number of positive and negative
pions (muons). The number of the last ones is (we choose $Z=7,A=14,L=15$):
\ba
N_p=M_p^2(\sigma_1^{QED}+\sigma_1^P)\approx M_p^2\sigma_1^P\approx 7.8,
\ea
where we can consider the only cross section of one pair or one $\sigma$ meson
with subsequent decay to the pair of charged pions.

Really the relative contribution
of two pairs with impact parameter exceeding $\rho>1 fm$ is suppressed by additional
factor $exp(-\rho m_\rho/2)|_{\rho\sim 1 fm}$.

Contribution of distant objects are suppressed by factor $exp(-R m_\rho/2)< 10^{-5}$
for $R>5fm$. So the importance of distant spectators become negligible.

It is known \cite{Adamson} that the ratio of positive charged muons to the
number of negative charged muons created in collision of the high energy
cosmic rays in the "knee" region of the spectrum $\Gamma=10^6$ with nuclei of
the atmospheric gas exceed unity:
\ba
R_{exp}=\frac{N_{\mu^+}}{N_{\mu^-}}=1.4\pm 0.03.
\ea

Our approach gives $R_{th}=1+(N_d/N_p)\approx 1.32$, in rather
satisfactory with $R_{exp}$. There are lot of mechanisms producing
muons in cosmic ray interaction with atmosphere\cite{Haungs}. Most
of them concern the production of soft muons not considered here.
\section{Conclusion}

Formulae to describe muon production can be applied to the energy range of LHC. The expected
cross section due to QED is $\sigma^{(1)}_{QED}=250 nb$; for Pomeron exchange mechanism
$\sigma^{(1)}_{QCD}=30\mu b$, so Pomeron mechanism dominate.

Cross sections of producing two and more muon pairs in one collision act is about 3 orders of
magnitude less than one pair production.

We do not consider mechanism of pair creation by proton fragmentation. Rough estimation give
at least two orders less contribution compared with the considered one.

As well we do not consider the mechanism with creation of
quark-antiquark creation in convolution of two reggeized gluons.
The relevant vertex being projected to $\sigma$ meson state turns
out to be at least an order of magnitude less than the one
considered. So it contribute negligible.

\section{Acknowledgements}
We are grateful to participants of BLTP JINR seminar for discussions.
Work was supported by grants INTAS 05-1000 008-8328 and
Byelorussian Foundation for Basic Research, grant No. F07D-005.

\end{document}